\documentclass[useAMS,usenatbib,usegraphicx]{mn2e}

\title[New Solutions for the Planetary Dynamics in HD160691 using a Newtonian Model and Latest Data]{New Solutions for the Planetary Dynamics in HD160691 using a Newtonian Model and Latest Data}
\author[D. Short]{D. Short$^{1}$, G. Windmiller$^{2}$\thanks{E-mail:
windmill@rohan.sdsu.edu}, and J. A. Orosz$^{3}$\\
$^{1}$Department of Mathematics and Statistics, San Diego State University, San Diego, CA 92182, USA\\
$^{2}$Department of Astronomy, San Diego State University, San Diego, CA 92182, USA\\
$^{3}$Department of Astronomy, San Diego State University, San Diego, CA 92182, USA}

\begin{document}

\date{Hopefully Accepted 2007; in original form 2007}

\pagerange{\pageref{firstpage}--\pageref{lastpage}} \pubyear{2006}

\maketitle

\label{firstpage}

\begin{abstract}
In this letter we present several new three and four-planet solutions based on the most current available radial velocity data for HD160691 ($\mu$Ara). These solutions are optimised using the Planetary Orbit Fitting Process (POFP) which is programmed and executed in MATLAB. POFP is based on a full integration of the system's multiple-body Newtonian equations of motion and on a multi level optimisation utilizing a variety of algorithms. The POFP solutions are presented in the context of the Keplerian-based solutions already appearing in the literature which we have reproduced here for comparison. The different solutions and their properties are compared over all data sets separately and combined. The new solutions do not seem to exhibit instabilities and are both co-planar and three-dimensional. We also provide a comparative prediction of the published and new solutions showing their diversion in the near future. In the short term, this projection will allow to choose between the variety of solutions as further observations are made.                                                                     
\end{abstract}


\section{Introduction}
In this paper we summarize our analysis to date of the planetary dynamics in the HD160691 ($\mu$Ara) system. Based on the most current data available in the literature we reproduce two of the most important published solutions and provide several new ones. The new solutions provided were achieved through our Planetary Orbit Fitting Process (POFP), which is an optimisation software written and operated in MATLAB, based on the work in Windmiller (2006) and Windmiller, Short $\&$ Orosz (2007). POFP provides multi-body solutions based on a full integration of the Newtonian equations of motion. Because of that, its solutions are general in their three-dimensional geometry and encompass all possible pathologies of the orbital dynamics. Optimising radial velocity (RV) data is an inverse problem, and therefore multiple solutions are expected to fit a single data set. In order to make a comparison between all solutions, previously-published Keplerian based solutions were translated into the same initial conditions used by the POFP scheme (i.e. the vectors of displacement and velocity in three dimensions at the given epoch) and reproduced in a setting allowing for gravitational interactions between all bodies in the system. We include a near future prediction showing the dispersion between both published and two of the new solutions, that will allow to determine the properties of the system as more data are collected.

\section{Background}
HD160691 is of spectral type G5V with $T_{\rm{eff}}=5807K$ and a stellar mass estimate of $1.08 M_{\sun}$ (McCarthy 2004). The Lick/Keck group reported in Butler et al. (2001), based on 2 years of observations, the existence of an exoplanet with a period of about 2 years. With an additional two years of data, Jones et al. (2003) confirmed the first planet and reported a linear trend. McCarthy et al. (2004) based on a total of 5.8 years of observations confirmed the existence of a second planet. Using the HARPS instrument, the European group Bouchy et al. (2006) mounted an astroseismologic campaign to study the stellar oscillations of HD160691. Bouchy et al. stated, ``The dispersion of each individual nights, in the range 1.5-2.5 $\rm{ms^{-1}}$, is strongly dominated by the acoustic modes with period around 8 minutes...''. In addition, they noted, ``...the presence of low frequency (few hours) modulations in the Doppler signal with semi-amplitudes of 1-2 $\rm{ms^{-1}}$ which are not averaged by 15 consecutive independent observations''. This provides a direct measurement of the nightly jitter for HD160691. These data along with additional measurements led to the claimed discovery of a small planet (minimum $14 M_{\earth}$) orbiting HD160691 with a period of 9.55 days by Santos (2004). Go$\acute{z}$dziewski et al. (2007), and Pepe et al. (2007), separately reported to have found a fourth planet. These last two are the 4 planet solutions we are augmenting here. 

\section{Comparing The Various Solutions}
The solutions generated by POFP are general in their geometry, and result from the direct integration of the Newtonian equations of motion. Because of that, they encompass all of the possible dynamical behavior caused by interaction between all bodies in the stellar-planetary system.  

The RV data sets we used for optimising solutions to the HD160691 system are those appearing up to this time in the literature: UCLES data (108 observations) in Butler et al. (2006), and CORALIE (40 observations) and HARPS (86 observations) data in Pepe et al. (2007). Two of the POFP solutions presented here are comparable in both RMS and RCS (reduced $\chi^{2}$) to those presented by Pepe et al. (2007) and by Go$\acute{z}$dziewski et al. (2007). The third exhibits over the entire data the lowest RMS and RCS to date. The latter is co-planar, as are the two published solutions. The formers serve as a reminder that the RV data in and of themselves do not predicate co-planarity. As in the cases of the published solutions, however, stability must be a requirement. POFP's output allows for a basic graphical examination of the orbit for stability behavior. After examining the orbit following an integration of $4 \times 10^{6}$ days we have deepened our search for instability. This was done by looking for excursions in both the eccentricity and instantaneous semi-major axis for each planet over the time span. We found no instabilities but those may occur over a much longer integration time.

In the presentation of their solution, Go$\acute{z}$dziewski et al. (2007) used a jitter\footnote{for a definition of stellar jitter and a discussion of this concept see, for example, Wright 2005} of 3.5 $\rm{ms^{-1}}$ (as reported in Butler 2006) and reported the parametric values valid at $T=2,451,118.89~[JD]$. Pepe quadratically augmented the error estimates for the HARPS data by 0.8 $\rm{ms^{-1}}$ (which includes the jitter and other factors) and reported the parametric values valid at $T=2,453,000~[JD]$. Any attempt to compare between these separate published solutions and those generated by POFP will require a common basis. Thus we have transformed the initial conditions of the published solutions to the POFP Instantaneous Keplerian Parameter (IKP) system and started the integration at the same epoch for all solutions.

The IKP parametrize a subset of all possible Newtonian initial conditions by requiring that these conditions for each planet instantaneously satisfy the Keplerian conditions for a bounded elliptical orbit. The IKP parameters are then given by the masses (of the star and planets) and, for the orbit of each planet, by $(x,y,z),\theta$ and $e$. These are each orbit's position of periastron, the ``delay angle'' from the periastron (comparable to the Keplerian orbital true anomaly), and eccentricity, respectively. In addition, the angular orientation of each orbital plane with respect to that of the first planet is used (two angles/rotations are computed from the planet's periastron coordinates and the rotation about the star-periastron axis, $\tau$, is given as an additional IKP parameter). The $n$-tuple of IKP is translated into Newtonian initial conditions for direct integration. The stellar velocity in three dimensions resulting from the integration is used to calculate the observer's direction from which this velocity best describes the RV data. This is done by using the singular value decomposition to optimise a least square solution for the `look vector' pointing to the observer. The result is put in terms of the inclination and the argument of periastron ($i,\omega$ respectively), of the first orbit. 

We note here that the published parameters are given to between 3 and 5 significant digits, thus limiting the accuracy of any reproduction of these solutions to this range of significant digits and limiting the RMS and RCS calculations to approximately 2 digits. We have chosen to use the planetary designations of Pepe et al. (2007). In addition, we quadratically subtracted 0.4 $\rm{ms^{-1}}$ of the 0.8 $\rm{ms^{-2}}$ augmentation for jitter of the HARPS error estimates made by them. We have used an estimate for the stellar jitter of 2 $\rm{ms^{-1}}$ (and 2.5 $\rm{ms^{-1}}$ for the 3-planet cases) based on the comment of Bouchy et al. cited above. We have also looked at the residuals from all of the solutions separately for the two more accurate data subsets, HARPS and UCLES, in both time and frequency domain, and find that an assumption of 2 $\rm{ms^{-1}}$ adequately describes the dispersion for these residuals with no evidence of further planets beyond the four.

\section{Tables and Figures}
The parameters of the published solutions reproduced by POFP at $T=2,451,118.89~[JD]$ and translated to standard Keplerian form are shown in Table 1. The slight change between these reproduced parameters and the ones actually published by Pepe et al. (2007) is due to planetary interaction during the time interval between our common initial condition time and their reporting time. The table continues with the parameters translated from the three 4-planet POFP solutions for the same epoch. The inclination found as a component of the optimal look-vector for each solution over the combined data set is: $i=~90,~90,~90,~172.55,~161.19 \rm{deg}$ for the Pepe, Go$\acute{z}$dziewski, POFP1, POFP2, and POFP3 respectively. The RCS and RMS values for the solutions, fitted against all data sets, are shown in Table 2.
 
Figure 1 shows the fit for the POFP1 solution with the residual velocities normally distributed. The orbital configuration of the POFP2 solution is shown in Figure 2 as seen through the observer's line of sight. The three dimensional shape of the orbit, shown in Figure 3 as the spread of points (each point is separated from the next by 1000 days) helps identify possible instabilities. The constancy of the orbits as well as the fact they don't cross is emphasized in Figure 4, showing the apastra and periastra of the three outer planets. The dispersion in the POFP1, POFP3 and the published solutions' near-future prediction is shown in figure 5.

\begin{table*}
 \centering
 \begin{minipage}{140mm}

  \caption{Solutions' parameters by planet}
  \begin{tabular}{@{}llccccccccc@{}}
  \hline\hline
& Planet & Time [JD]\footnote{This is the epoch for which the other parameters hold. These parameters are, in order: Period, Time of Periastron passage, Mean Longitude at $T_{0}$, Eccentricity, Longitude of Periastron, Mean Motion, Semi-major Axis, and planetary Mass}
 & P [days]   & $T_{0}$ [JD] \footnote{[JD] - 2450000} & $\lambda$ [deg] & e     & $\omega$ [deg]  & K $[\rm{ms^{-1}}]$ & a [AU] & Mass $[M_{J}]$\\
\hline
 Pepe\footnote{The first line is the solution for planet b that was published by Pepe et al. (2007). All other lines by either Pepe or Go$\acute{z}$dziewski are reproductions in POFP after translation from the original Keplerian scheme of Pepe et al. (2007) and Go$\acute{z}$dziewski et al. (2007). The change of parameters between the first and second lines attests to planetary gravitational interaction.}

                         & b      & 2453000   & ~~643.25   & 2365.60      & ~17.60          & 0.128 & ~~22.00         & 37.780        & 1.497  & -- \\
 Pepe                    & b      & 2451119   & ~~646.11   & 1075.97      & ~49.01          & 0.132 & ~~25.09         & 37.688        & 1.501  & 1.675 \\
 Go$\acute{z}$dziewski   & b      & 2451119   & ~~646.49   & 2721.80      & ~49.70          & 0.000 & 222.20          & 35.870        & 1.534  & 1.677 \\
 POFP 1                  & b      & 2451119   & ~~643.89   & 1112.95      & ~45.81          & 0.024 & ~42.49          & 35.994        & 1.498  & 1.611 \\
 POFP 2                  & b      & 2451119   & ~~642.55   & ~856.72      & ~42.51          & 0.151 & 255.63          & 36.070        & 1.474  & 3.265 \\
 POFP 3                  & b      & 2451119   & ~~646.45   & 1388.66      & ~47.35          & 0.078 & 197.59          & 36.359        & 1.503  & 5.070 \\
 Pepe                    & c      & 2451119   & ~~~~~9.64   & 1121.21     & 127.82          & 0.172 & 214.63          & ~3.060        & 0.091  & 0.033 \\
 Go$\acute{z}$dziewski   & c      & 2451119   & ~~~~~9.64   & ~632.60     & 124.40          & 0.184 & 313.30          & ~2.830        & 0.093  & 0.032 \\
 POFP 1                  & c      & 2451119   & ~~~~~9.64   & 1119.01     & 107.74          & 0.132 & 112.24          & ~2.897        & 0.091  & 0.032 \\
 POFP 2                  & c      & 2451119   & ~~~~~9.64   & 1118.87     & 157.48          & 0.150 & 156.90          & ~3.074        & 0.090  & 0.256 \\
 POFP 3                  & c      & 2451119   & ~~~~~9.64   & 1122.82     & 120.88          & 0.050 & 267.94          & ~2.699        & 0.091  & 0.093 \\
 Pepe                    & d      & 2451119   & ~~309.61   & 1171.21      & 144.37          & 0.049 & 205.21          & 14.903        & 0.919  & 0.522 \\
 Go$\acute{z}$dziewski   & d      & 2451119   & ~~307.48   & 3070.40      & 127.10          & 0.079 & 251.80          & 13.190        & 0.934  & 0.480 \\
 POFP 1                  & d      & 2451119   & ~~308.59   & 1160.36      & 125.66          & 0.061 & 174.04          & 12.239        & 0.917  & 0.428 \\
 POFP 2                  & d      & 2451119   & ~~299.63   & 1234.12      & ~76.19          & 0.048 & 214.64          & ~8.501        & 0.886  & 2.251 \\
 POFP 3                  & d      & 2451119   & ~~308.58   & 1146.26      & 117.67          & 0.060 & 149.60          & ~9.560        & 0.917  & 1.042 \\
 Pepe                    & e      & 2451119   & ~4129.10   & 2954.01      & 260.63          & 0.097 & ~60.63          & 21.905        & 5.171  & ~1.814 \\
 Go$\acute{z}$dziewski   & e      & 2451119   & ~4440.79   & 4149.30      & 267.60          & 0.027 & 153.30          & 27.250        & 5.543  & ~2.423 \\
 POFP 1                  & e      & 2451119   & ~3403.04   & ~411.91      & 236.57          & 0.059 & 161.78          & 20.371        & 4.545  & ~1.586 \\
 POFP 2                  & e      & 2451119   & ~3410.28   & -321.04      & 235.16          & 0.015 & ~83.16          & 20.924        & 4.484  & ~1.587 \\
 POFP 3                  & e      & 2451119   & ~6611.59   & 1773.78      & 292.93          & 0.095 & 328.59          & 37.566        & 7.094  & ~9.941 \\ 
\hline
\end{tabular}
\end{minipage}
\end{table*}

\begin{table*}
 \centering
 \begin{minipage}{140mm}

  \caption{Fitness values - RCS ($\chi_{\nu}^{2}$) and RMS - for all solutions}
  \begin{tabular}{@{}lccccccccc@{}}
  \hline\hline
 Data Set & Combined & Combined & Combined & HARPS & HARPS & HARPS & UCLES & UCLES & UCLES\\ 
 Fitness\footnote{ The number of degrees of freedom for the 3-planet cases is 14 for the reproduced solutions and POFP1, 18 for POFP2 and POFP3. For the 4-planet cases it is 19 for the reproduced solutions and POFP1, 25 for POFP2 and POFP3. The number of points for the Combined, HARPS, and UCLES data sets are 234, 86, and 108 respectively}  
 & RCS & CS & RMS\footnote{in $\rm{ms^{-1}}$} & RCS & CS & RMS & RCS & CS & RMS\\
 Solution:\footnote{The jitter for the 3-planet cases is 2.5 $\rm{ms^{-1}}$ and for the 4-planet cases is 2.0 $\rm{ms^{-1}}$. 0.4 $\rm{ms^{-1}}$ has been quadratically subtracted from the HARPS data.}\\
\hline
Pepe\\
~~3-planet & 1.55 & 339.6 & 4.14 & 1.18 & 83.7 & 2.57 & 1.60 & 148.4 & 3.69\\
~~4-planet & 1.45 & 309.8 & 3.73 & 0.63 & 41.5 & 1.49 & 1.65 & 145.0 & 3.18\\
~~\%Change\footnote{Improvement in fit from the 3 to the 4-planet solution by adding a small short period planet} 
           &      &       & 10\% &      &      &42\% &       &       &14\%\\
Go$\acute{z}$dziewski et al.\\
~~3-planet & 1.64 & 360.2 & 4.13 & 2.29 & 162.8 & 3.56 & 1.09 & 101.0 & 3.02\\
~~4-planet & 1.61 & 344.1 & 3.65 & 2.52 & 166.6 & 2.93 & 0.88 & ~77.3 & 2.32\\
~~\%Change &      &       &12\%  &      &       &18\% &       &       &23\%\\
POFP 1\\
~~3-planet & 1.40 & 306.9 & 3.88 & 1.39 & 98.9 & 2.79 & 1.22 & 113.5 & 3.16\\
~~4-planet & 1.26 & 270.7 & 3.42 & 1.03 & 68.3 & 1.91 & 1.12 & ~98.9 & 2.58\\
~~\%Change &      &       &12\%  &      &       &32\% &      &       &18\%\\
POFP 2\\
~~3-planet & 1.40 & 306.9 & 3.88 & 1.39 & 98.9 & 2.79 & 1.22 & 113.5 & 3.16\\
~~4-planet & 1.63 & 339.8 & 3.75 & 1.32 & 79.1 & 2.05 & 1.96 & 160.6 & 3.28\\
~~\%Change &      &       & 9\% &       &      &35\% &       &       & 7\%\\
POFP 3\\
~~3-planet & 1.59 & 341.3 & 4.14 & 1.49 & 99.9 & 2.81 & 1.57 & 139.9 & 3.57\\
~~4-planet & 1.43 & 298.1 & 3.59 & 1.29 & 77.6 & 2.03 & 1.46 & 119.8 & 2.89\\
~~\%Change &      &       &13\% &       &      &28\% &       &       &19\%\\
\hline
\end{tabular}
\end{minipage}
\end{table*}

\begin{figure}
 \includegraphics[width=80mm]{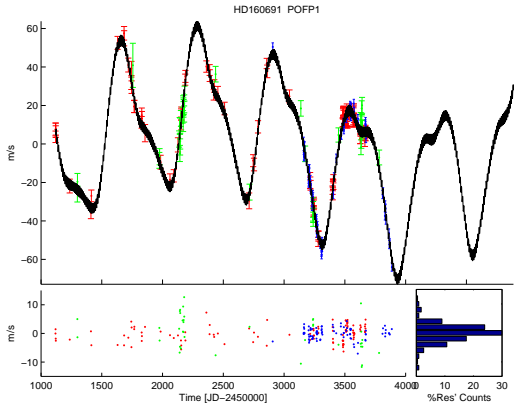}
 \caption{The RV fit for the POFP1 solution (see Tables 1,2). The different data sets are designated by colour: UCLES in red, HARPS in blue and CORALIE in green. Note the histogram showing the distribution of the residual velocities, which makes a reasonable Gausian.}
\end{figure}

\begin{figure}
 \includegraphics[width=80mm]{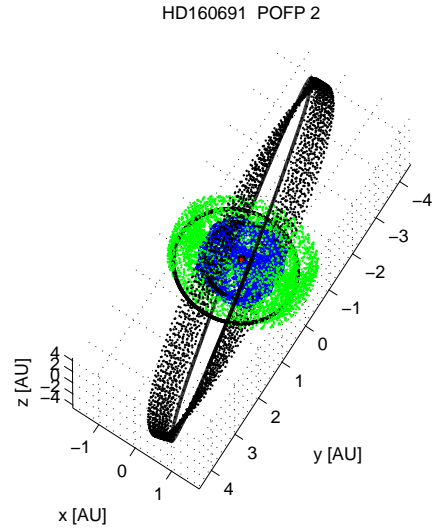}
 \caption{The orbital configuration around the system's centre of mass for the POFP2 solution. The dark lines represent the first full orbit. The points represent the location of each planet in 1000 days intervals (planets b,c,d,e are the green,blue,red,black respectively). The overall time of integration shown is $4 \times 10^{6}$ days. The view angle nears the observer's line of site (down the look-vector)}.
\end{figure}

\begin{figure}
 \includegraphics[width=80mm]{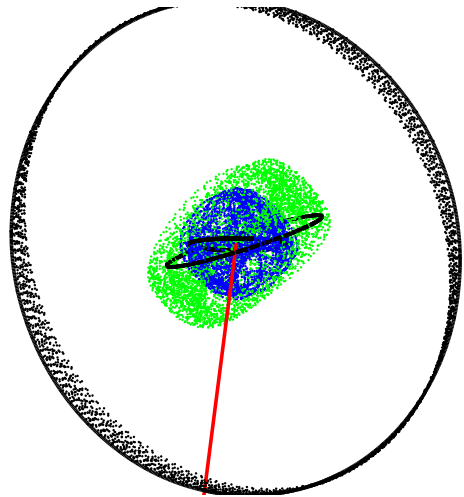}
 \caption{The Orbital configuration around the system's centre of mass for the POFP2 solution from a different angle. Graphical Inspection of the spread of points from convenient points of view can help identify instabilities. The red arrow at the centre is the look-vector, pointing in the observer's line of sight.}
\end{figure}

\begin{figure}
 \includegraphics[width=80mm]{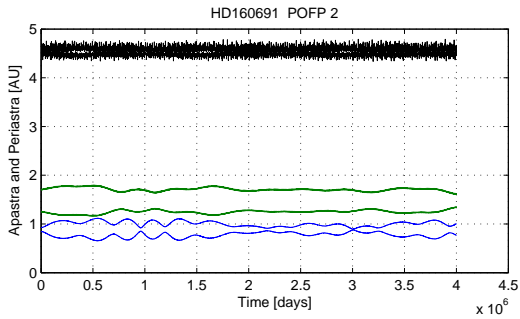}
 \caption{The Apastra and Periastra of the three outer planets in the POFP2 solution, during the $4 \times 10^{6}$ days integration time span. The planetary orbits do not cross.}
\end{figure}

\begin{figure}
 \includegraphics[width=80mm]{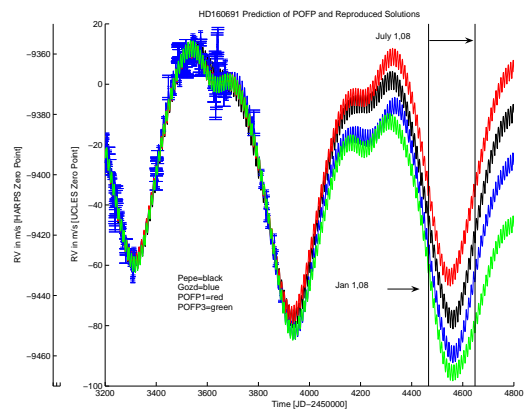}
 \caption{Projection of two reproduced published solutions along with POFP1 and POFP3 into the near future. Note that data taken at the present and during 2008 can already help distinguish the more viable solutions.}
\end{figure}

\bsp

\label{lastpage}

\end{document}